\documentclass[12pt]{article}

\catcode`@=12
\begin{document}
\thispagestyle{empty}
\begin{flushright} 
UCRHEP-T334\\ 
April 2002\
\end{flushright}
\vspace{0.5in}
\begin{center}
{\LARGE	\bf Neutrino Mass from Triplet and\\ Doublet Scalars at the TeV 
Scale\\}
\vspace{1.5in}
{\bf Ernest Ma\\}
\vspace{0.2in}
{\sl Physics Department, University of California, Riverside, 
California 92521\\}
\vspace{1.5in}
\end{center}
\begin{abstract}\
If the minimal standard model of particle interactions is extended to include 
a scalar triplet with lepton number $L=-2$ and a scalar doublet with $L=-1$, 
neutrino masses $m_\nu \sim \mu_{12}^4 v^2/M^5 \sim 10^{-2}$ eV is possible, 
where $v \sim 10^2$ GeV is the electroweak symmetry breaking scale, $M \sim 1$ 
TeV is the typical mass of the new scalars, and $\mu_{12} \sim 1$ GeV is a 
soft lepton-number-violating parameter.
\end{abstract}

\newpage
\baselineskip 24pt

In the minimal standard model of particle interactions, neutrinos are 
massless, but if new interactions exist at a higher scale, then they may 
become massive through the unique effective dimension-five operator 
\cite{wein}
\begin{equation}
{\cal L}_{eff} = {f_{ij} \over \Lambda} L_i L_j \Phi \Phi,
\end{equation}
where $L_i = (\nu_i,l_i)_L$ is the usual left-handed lepton doublet, 
$\Phi = (\phi^+,\phi^0)$ is the usual scalar Higgs doublet, and $\Lambda$ 
is an effective large mass.  There are three tree-level realizations 
\cite{ma98} of this operator: (I) the canonical seesaw mechanism \cite{seesaw} 
using one heavy right-handed neutrino $N_R$ for each $\nu_i$; (II) the 
addition of a heavy Higgs triplet $(\xi^{++}, \xi^+, \xi^0)$ which couples to 
$L_i L_j$ directly \cite{scva,masa}; and (III) the replacement of $N_R$ with 
$(\Sigma^+, \Sigma^0, \Sigma^-)_R$ \cite{flhj,ma02}.  If the mass of $N$ or 
$\xi$ or $\Sigma$ is very large, then each realization is the same as any 
other at low energies, because the only observable effect would be the 
appearance of small Majorana neutrino masses.

It has recently been pointed out that simple extensions of the above 
minimal scenarios for neutrino mass are possible for which the scale of 
new physics may be only a few TeV and thus be observable at 
future accelerators.  There are two specific proposals: (A) the Higgs 
triplet $\xi$ may be only a few TeV, whose decay into two leptons would 
map out all elements of the neutrino mass matrix \cite{marasa}; and (B) 
the fermion singlets $N_R$ as well as a second Higgs doublet $\eta = 
(\eta^+,\eta^0)$ may be only a few TeV \cite{ma01}.

In (A), the notion of lepton-number violation as a distance effect from 
the separation of our brane from another in the context of large extra 
dimensions is invoked to explain the smallness of the trilinear scalar 
coupling of $\xi$ to $\Phi \Phi$.  In (B), there is no need to consider 
large extra dimensions.  Instead, the effective operator of Eq.~(1) is 
suppressed because $\Phi$ is replaced with $\eta$, which has a naturally 
small vacuum expectation value.  In this note, the two mechanisms are 
synthesized so that there is a Higgs triplet $\xi$ from (A), and a Higgs 
doublet $\eta$ from (B), but no $N_R$.  Neutrino masses come from $\xi$ 
(which is assigned lepton number $L=-2$), and its interaction with $\eta$ 
(which has $L=-1$).  The smallness of $m_\nu$ comes from the soft breaking 
of $L$ in the scalar sector, with the result
\begin{equation}
m_\nu \sim {\mu_{12}^4 v^2 \over M^5} \sim 10^{-2} ~{\rm eV},
\end{equation}
where $v \sim 10^2$ GeV is the electroweak symmetry breaking scale, $M \sim 1$ 
TeV is the typical mass of the new scalars, and $\mu_{12} \sim 1$ GeV is a 
soft lepton-number-violating parameter.  This model requires neither large 
extra dimensions nor $N_R$ to obtain naturally small Majorana neutrino 
masses, and have verifiable experimental consequences at the TeV scale.

In the Higgs triplet model, $\xi$ couples to leptons according to
\begin{equation}
{\cal L}_Y = f_{ij} [\xi^0 \nu_i \nu_j + \xi^+ (\nu_i l_j + l_i \nu_j)/\sqrt 2 
+ \xi^{++} l_i l_j] + H.c.,
\end{equation}
resulting in $(m_\nu)_{ij} = 2 f_{ij} \langle \xi^0 \rangle$.  Therefore, 
in order to explain why neutrino masses are so small, a natural mechanism 
for $\langle \xi^0 \rangle$ to be small is needed.  Consider now 
the Higgs sector consisting of the usual standard-model doublet $\Phi$ 
(with $L=0$), a second doublet $\eta$ (with $L=-1$), and a triplet $\xi$ 
(with $L=-2$).  Let
\begin{equation}
\Delta \equiv \left( \begin{array} {c@{\quad}c} \xi^+/\sqrt 2 & \xi^{++} \\ 
\xi^0 & -\xi^+/\sqrt 2 \end{array} \right),
\end{equation}
then the most general $L$-conserving Higgs potential is given by
\begin{eqnarray}
V &=& m_1^2 \Phi^\dagger \Phi + m_2^2 \eta^\dagger \eta + m_3^2 
Tr \Delta^\dagger \Delta \nonumber \\ &+& {1 \over 2} \lambda_1 (\Phi^\dagger 
\Phi)^2 + {1 \over 2} \lambda_2 (\eta^\dagger \eta)^2 + {1 \over 2} \lambda_3 
\left( Tr \Delta^\dagger \Delta \right)^2 + {1 \over 2} \lambda_4 \left( 
Tr \Delta^\dagger \Delta^\dagger \right) \left( Tr \Delta \Delta \right) 
\nonumber \\ &+& \lambda_5 (\Phi^\dagger \Phi)(\eta^\dagger \eta) + 
\lambda_6 (\Phi^\dagger \Phi) \left( Tr \Delta^\dagger \Delta 
\right) + \lambda_7 (\eta^\dagger \eta) \left( Tr \Delta^\dagger 
\Delta \right) \nonumber \\ &+& \lambda_8 (\Phi^\dagger \eta)(\eta^\dagger 
\Phi) + \lambda_9 \left( \Phi^\dagger \Delta^\dagger \Delta \Phi \right) + 
\lambda_{10} \left( \eta^\dagger \Delta^\dagger \Delta \eta \right) \nonumber 
\\ &+& \mu \left( \eta^\dagger \Delta \tilde \eta \right) + H.c.,
\end{eqnarray}
where $\tilde \eta = (\bar \eta^0, -\eta^-)$ and the parameter $\mu$ has the 
dimension of mass.  Lepton number is then assumed to be broken by explicit 
$soft$ terms, i.e.
\begin{equation}
V_{soft} = \mu_{12}^2 \Phi^\dagger \eta + \mu' \left( \Phi^\dagger \Delta 
\tilde \eta \right) + \mu'' \left( \Phi^\dagger \Delta \tilde \Phi \right) 
+ H.c.
\end{equation}

Let $\langle \phi^0 \rangle = v_1$, $\langle \eta^0 \rangle = v_2$, and 
$\langle \xi^0 \rangle = v_3$, then the minimum of $V$ is given by
\begin{eqnarray}
V_{min} &=& m_1^2 v_1^2 + m_2^2 v_2^2 + m_3^2 v_3^2 + 2 \mu_{12}^2 v_1 v_2 
\nonumber \\ &+& {1 \over 2} \lambda_1 v_1^4 + {1 \over 2} \lambda_2 v_2^4 
+ {1 \over 2} \lambda_3 v_3^4 + (\lambda_5 + \lambda_8) v_1^2 v_2^2 + 
\lambda_6 v_1^2 v_3^2 + \lambda_7 v_2^2 v_3^2 \nonumber \\ &+& 2 \mu v_2^2 v_3 
+ 2 \mu' v_1 v_2 v_3 + 2 \mu'' v_1^2 v_3,
\end{eqnarray}
where $\mu_{12}^2$, $\mu$, $\mu'$, and $\mu''$ have been assumed real for 
simplicity.  The equations of constraint are obtained from $\partial V_{min} 
/\partial v_i = 0$, i.e.
\begin{eqnarray}
v_1[m_1^2 + \lambda_1 v_1^2 + (\lambda_5+\lambda_8) v_2^2 + \lambda_6 v_3^2 
+ 2 \mu'' v_3] + \mu_{12}^2 v_2 + \mu' v_2 v_3 &=& 0, \\ 
v_2[m_2^2 + \lambda_2 v_2^2 + (\lambda_5 + \lambda_8) v_1^2 + \lambda_7 v_3^2 
+ 2 \mu v_3] + \mu_{12}^2 v_1 + \mu' v_1 v_3 &=& 0, \\ 
v_3[m_3^2 + \lambda_3 v_3^2 + \lambda_6 v_1^2 + \lambda_7 v_2^2] + \mu v_2^2 
+ \mu' v_1 v_2 + \mu'' v_1^2 &=& 0.
\end{eqnarray}
Consider now the case $m_1^2 < 0$, but $m_2^2 > 0$ and $m_3^2 > 0$ with small 
$\mu_{12}^2$, $\mu'$ and $\mu''$.  The solutions to the above equations are 
then
\begin{eqnarray}
v_1^2 &\simeq& - {m_1^2 \over \lambda_1}, \\ v_2 &\simeq& - {\mu_{12}^2 v_1 
\over m_2^2 + (\lambda_5 + \lambda_8) v_1^2}, \\ v_3 &\simeq& - {\mu v_2^2 + 
\mu' v_1 v_2 + \mu'' v_1^2 \over m_3^2 + \lambda_6 v_1^2}.
\end{eqnarray}
Since $\mu'$ violates $L$ by 1 unit and $\mu''$ violates $L$ by 2 units, it 
is reasonable to assume that $\mu''/\mu' \sim \mu'/\mu \sim v_2/v_1$.  Thus 
for $m_2$, $m_3$, and $\mu$ all approximated by $M \sim 1$ TeV,
\begin{equation}
v_2 \sim {\mu_{12}^2 v_1 \over M^2}, ~~~ v_3 \sim {v_2^2 \over M}.
\end{equation}
This shows that $v_3 << v_2 << v_1$, and
\begin{equation}
v_3 \sim {\mu_{12}^4 v_1^2 \over M^5},
\end{equation}
i.e. the analog of Eq.~(2).  For $v_1 \sim 10^2$ GeV and $\mu_{12} \sim 1$ 
GeV, the solutions are $v_2 \sim 0.1$ MeV and $v_3 \sim 10^{-2}$ eV as 
desired.

In conclusion, a scenario has been presented where a Higgs triplet at the TeV 
scale is responsible for neutrino masses.  The decay of $\xi^{++}$ into 
$l^+_i l^+_j$ would then map out the neutrino mass matrix, as proposed 
previously \cite{marasa}.  In addition, a second Higgs doublet at the TeV 
scale is predicted \cite{ma01} so that $\xi^{++} \to \eta^+ \eta^+$ is also 
possible if kinematically allowed.\\[5pt]

This work was supported in part by the U.~S.~Department of Energy under 
Grant No.~DE-FG03-94ER40837.

\newpage
\bibliographystyle{unsrt}

\begin{thebibliography}{99}
\bibitem{wein} S. Weinberg, Phys. Rev. Lett. {\bf 43}, 1566 (1979).
\bibitem{ma98} E. Ma, Phys. Rev. Lett. {\bf 81}, 1171 (1998).
\bibitem{seesaw} M. Gell-Mann, P. Ramond, and R. Slansky, in 
{\em Supergravity}, edited by P. van Nieuwenhuizen and D. Z. Freedman 
(North-Holland, Amsterdam, 1979), p.~315; T. Yanagida, in {\em Proceedings 
of the Workshop on the Unified Theory and the Baryon Number in the Universe}, 
edited by O. Sawada and A. Sugamoto (KEK Report No.~79-18, Tsukuba, Japan, 
1979), p.~95; R. N. Mohapatra and G. Senjanovic, Phys. Rev. Lett. {\bf 44}, 
912 (1980).
\bibitem{scva} J. Schechter and J. W. F. Valle, Phys. Rev. {\bf D22}, 2227 
(1980).
\bibitem{masa} E. Ma and U. Sarkar, Phys. Rev. Lett. {\bf 80}, 5716 (1998); 
T. Hambye, E. Ma, and U. Sarkar, Nucl. Phys. {\bf B602}, 23 (2001).
\bibitem{flhj} R. Foot, H. Lew, X.-G. He, and G. C. Joshi, Z. Phys. {\bf C44}, 
441 (1989).
\bibitem{ma02} E. Ma, hep-ph/0112232 (Mod. Phys. Lett. {\bf A}, in press).
\bibitem{marasa} E. Ma, M. Raidal, and U. Sarkar, Phys. Rev. Lett. {\bf 85}, 
3769 (2000); Nucl. Phys. {\bf B615}, 313 (2001).
\bibitem{ma01} E. Ma, Phys. Rev. Lett. {\bf 86}, 2502 (2001).
\end{thebibliography}

\end{document}